\documentclass[amsmath,amssymb,floatfix,preprintnumbers,nofootinbib,prd,twocolumn,superscriptaddress]{revtex4}

\usepackage[utf8]{inputenc}
\usepackage{graphicx}

\usepackage[format=plain,justification=raggedright, singlelinecheck=false, font=small]{caption}
\usepackage[format=plain,justification=raggedright, singlelinecheck=false,font=small]{subcaption}

\usepackage{amssymb}
\usepackage{amsxtra}
\usepackage{amsmath}
\usepackage{booktabs,multirow,tabularx}
\usepackage{slashed}
\usepackage{float}
\usepackage{placeins}
\usepackage{rotating}
\usepackage{lscape}
\usepackage{color}
\usepackage{hyperref}
\usepackage{cancel}
\usepackage[Q=yes,pverb-linebreak=no]{examplep}
\usepackage{ulem}

\DeclareGraphicsExtensions{.pdf}
\graphicspath{{./figures/}}

\newcommand{\lsim}{\lesssim}
\newcommand{\gsim}{\gtrsim}

\newcommand{\ord}[1]{\mathcal{O}{(#1)}}

\def\beq{\begin{equation}}
\def\bea{\begin{eqnarray}}
\def\eeq{\end{equation}}
\def\eea{\end{eqnarray}}
\def\beqal{\begin{align}}
\def\endal{\end{align}}

\definecolor{applegreen}{rgb}{0.55, 0.71, 0.0}
\definecolor{purple}{rgb}{0.5,0.0,0.5}

\makeatletter
\newcommand\footnoteref[1]{\protected@xdef\@thefnmark{\ref{#1}}\@footnotemark}
\makeatother

\DeclareFontFamily{U}{cbgreek}{}
\DeclareFontShape{U}{cbgreek}{m}{n}{
        <-6>    grmn0500
        <6-7>   grmn0600
        <7-8>   grmn0700
        <8-9>   grmn0800
        <9-10>  grmn0900
        <10-12> grmn1000
        <12-17> grmn1200
        <17->   grmn1728
      }{}
\DeclareFontShape{U}{cbgreek}{bx}{n}{
        <-6>    grxn0500
        <6-7>   grxn0600
        <7-8>   grxn0700
        <8-9>   grxn0800
        <9-10>  grxn0900
        <10-12> grxn1000
        <12-17> grxn1200
        <17->   grxn1728
      }{}

\makeatletter
\newcommand{\normalorbold}{%
  \ifnum\pdf@strcmp{\math@version}{bold}=\z@ bx\else m\fi
}
\makeatother

\begin{document}

\title{\boldmath Flavor-Violating ALPs, Electron $g-2$, and the Electron-Ion Collider}

\author{Hooman Davoudiasl\footnote{email: hooman@bnl.gov}}

\affiliation{High Energy Theory Group, Physics Department,Brookhaven National Laboratory, Upton, NY 11973, USA}

\author{Roman Marcarelli\footnote{email: roman.marcarelli@colorado.edu}}

\affiliation{High Energy Theory Group, Physics Department,Brookhaven National Laboratory, Upton, NY 11973, USA}

\affiliation{Department of Physics, University of Colorado, Boulder, Colorado 80309, USA}

\author{Ethan T. Neil\footnote{email: ethan.neil@colorado.edu}}

\affiliation{Department of Physics, University of Colorado, Boulder, Colorado 80309, USA}

\begin{abstract}

We revisit the possibility that light axion-like particles (ALPs) with lepton flavor violating couplings could give significant contributions to the electron's anomalous magnetic moment $g_e-2$.  Unlike flavor diagonal lepton-ALP couplings, which are exclusively axial, lepton flavor violating couplings can have arbitrary chirality.  Focusing on the $e$-$\tau$ ALP coupling, we find that the size of the contribution to $g_e-2$ depends strongly on the chirality of the coupling.  A significant part of the parameter space for which such a coupling can explain experimental anomalies in $g_e-2$ can be probed at the Electron-Ion Collider, which is uniquely sensitive to the chirality of the coupling using the polarization of the electron beam.  

\end{abstract}

\maketitle

\section{Introduction\label{sec:intro}}

Axion-like particles (ALPs) are a generic type of new particle that can occur in a wide variety of scenarios for physics beyond the Standard Model (BSM).  Such particles commonly arise as pseudo-Nambu-Goldstone bosons associated with new physics \cite{Jaeckel:2010ni,Irastorza:2018dyq}.  Their pseudo-Goldstone nature means that ALPs can naturally be very light compared to the scale of new physics, with very weak interactions suppressed by the same large mass  scale.

Within the general theory space, an intriguing possibility is the presence of significant lepton-flavor-violating (LFV) couplings for an ALP.  Lepton flavor is almost perfectly conserved in the Standard Model (SM), up to neutrino flavor mixing, so that lepton flavor violation is a sensitive probe of new physics.  At the same time, existing anomalies in precision lepton physics, in particular in the anomalous magnetic moments of the electron and muon, motivate the exploration of new particles coupled to leptons.  Previous studies of the phenomenology of LFV ALPs include \cite{Gelmini:1982zz,Tsumura:2009yf,Bauer:2019gfk,Cornella:2019uxs,Endo:2020mev,Davoudiasl:2021haa,Davoudiasl:2021mjy,Bauer:2021mvw,Cheung:2021mol,Jho:2022snj,Knapen:2023zgi}, and searches for ALPs at the EIC include  \cite{Liu:2021lan,Balkin:2023gya}.

While there is significant attention on present tensions between theory and experiment in the anomalous magnetic moment of the muon \cite{Aoyama:2020ynm}, the anomalous magnetic moment of the electron $g_e-2$ is also showing possible tensions with the SM, although the situation is somewhat less clear.  The latest measurement of $g_e-2$ \cite{Fan:2022eto} is in tension with the SM  prediction of either $+2.2\sigma$ or $-3.7\sigma$ \cite{Davoudiasl:2023huk}, depending on whether the fine-structure constant $\alpha$ obtained from Rb \cite{Morel:2020dww} or Cs \cite{Parker:2018vye} is used as input.

Contributions from lepton-flavor violating ALP couplings to $g_e-2$ have been calculated and studied in the literature previously \cite{Bauer:2019gfk,Cornella:2019uxs,Bauer:2021mvw}.  In this work, we focus on the interplay between the magnitude of the $e$-$\tau$ ALP coupling and its chirality, which has a significant impact on the $g_e-2$ contribution.  We find a substantial region of parameter space where a single $e$-$\tau$ coupling with a GeV-scale ALP and the correct chirality can explain the current experimental anomalies in $g_e-2$ without being constrained by other experiments, but with discovery potential in future searches at the EIC \cite{Davoudiasl:2021mjy}.  The EIC is also uniquely well-equipped to probe the chiral structure of such a coupling directly by utilizing the polarization of its electron beam.
 
\section{Model Setup\label{sec:model}}

We will consider an ALP-$e$-$\tau$ interaction of the form
\begin{equation}
{\cal L}_{\rm int} = \frac{\partial_\mu a}{\Lambda}\bar{\tau}\gamma^\mu\left(V_{\tau e} + A_{\tau e}\gamma_5\right)e + \text{\small H.C.},
\end{equation}
where $V_{\tau e}$ and $A_{\tau e}$ are complex constants in general. We can decompose this term into a more useful form by defining \begin{align}
    \phi_V & \equiv \arg{V_{\tau e}},  & \phi_A &\equiv \arg{A_{\tau e}} \nonumber\\
    \phi &\equiv \phi_A - \phi_V, &
 \tan{\theta} &\equiv  \left(|V_{\tau e}|/|A_{\tau e}|\right),\\
  C_{\tau e} &\equiv \sqrt{|A_{\tau e}|^2 + |V_{\tau e}|^2}. \nonumber
\end{align}
Then, the interaction can be written as 
\begin{equation}
    {\cal L}_{\rm int} = \frac{C_{\tau e}e^{i\phi_V}}{\Lambda}\partial_\mu a \, \bar{\tau}\gamma^\mu\left(\sin{\theta} + e^{i\phi}\cos{\theta} \gamma_5\right)e + \text{\small H.C.}
    \label{eq:Lint}
\end{equation}
Note that if the coupling were lepton flavor diagonal, hermiticity would require $\theta = \phi_A = \phi_V  = 0$. Hence, the presence of flavor violation allows for both parity (P) violation (via  $\theta$) and charge/parity (CP) violation (via $\phi$ and $\phi_V$).  The overall phase $\phi_V$ does not contribute to any of the physical processes we consider below.

For simplicity, we will study the case in which the only significant coupling is $C_{\tau e}$.  As explored in Ref.~\cite{Davoudiasl:2021haa}, other bounds from precision flavor physics are relevant for ALPs heavier than $m_\tau$ only when lepton-flavor diagonal couplings are also present and sufficiently strong.

\section{Contributions to electron EDM and \boldmath$g-2$ \label{sec:electron-moments}}

LFV ALPs which couple to electrons and $\tau$ leptons can contribute to the electric and magnetic dipole moments of the electron. The electric and magnetic dipole moment can be extracted from the electron-photon interaction vertex $\Gamma^\mu$,  which can be decomposed into the general form \cite{Nowakowski:2004cv}
\bea
    \Gamma^\mu &=& \gamma^\mu F_1(q^2) + \frac{i\sigma^{\mu\nu}}{2m_e}q_\nu F_2(q^2) + \frac{i\sigma^{\mu\nu}}{2m_e}q_\nu  \gamma_5 F_3(q^2)  \nonumber\\
    &+&\frac{1}{2m_e}\left(q^\mu - \frac{q^2}{2m_e}\gamma^\mu\right)\gamma_5 F_4(q^2)\,,
\eea
where $\sigma^{\mu\nu} = \frac{i}{2} \left[\gamma^\mu, \gamma^\nu\right]$, $q=p_2-p_1$ is the external photon momentum, $p_1$ is the initial momentum of the electron, $p_2$ is the final momentum of the electron; for an on-shell amplitude $q^2 = 0$.  The form factors $F_i$ are all real \cite{Nowakowski:2004cv}. Each form factor corresponds to a property of the electron: $F_1$ yields the electric charge, $F_2$ the magnetic dipole moment, $F_3$ the electric dipole moment (EDM), and $F_4$ the anapole moment. In particular, the electric and magnetic moments are given by
\begin{align}
    a_e = \frac{g_e-2}{2} =  F_2(0); & & d_e = -\frac{e}{2m_e}F_3(0).
\end{align}

In the limit $m_\tau \gg m_e$, the contributions to these dipole moments from the interaction (\ref{eq:Lint}) are
\begin{align}
\Delta a_e &=  -\frac{m_e^2 C_{\tau e}^2}{16\pi^2\Lambda^2}\left(f(x_\tau) + \frac{m_\tau}{m_e}g(x_\tau)\cos{2\theta}\right), \label{eq:mdm}\\
\Delta d_e &= -\frac{e}{2m_e}\frac{m_\tau m_e C_{\tau e}^2}{{16\pi^2 \Lambda^2}}g(x_\tau)\sin{2\theta}\sin{\phi}\,,\label{eq:edm}
\end{align}
where $x_\tau \equiv m_a^2 / m_\tau^2$ and $f$ and $g$ are given by
\begin{align}
    f(x) &= \frac{2x^2(2x-1)}{(x-1)^4}\log{x} - \frac{5-19x+20x^2}{3(x-1)^3},\\
    g(x) &= \frac{2x^2}{(x-1)^3}\log{x} + \frac{1-3x}{(x-1)^2}.
\end{align}
These results were obtained using PackageX \cite{Patel:2016fam} in Mathematica and are in agreement with Refs.~\cite{Cornella:2019uxs,Bauer:2019gfk} in the appropriate limits. Our result for $\Delta d_e$ reproduces the formula from Ref.~\cite{DiLuzio:2020oah} for the contribution from flavor-violating lepton-ALP couplings.

The first term in the contribution to the magnetic dipole moment (\ref{eq:mdm}) is often neglected for $m_\tau \gg m_e$, but we emphasize that this is not the case for chiral interactions, where $\theta = \pm\pi/4$. In particular, we find that for chiral or near-chiral interactions, the form factor $F_2$ is suppressed by $m_e/m_\tau$ but is still non-zero. The electric dipole moment, on the other hand, is proportional to $\sin{2\theta}\sin{\phi}$, so it is only non-zero when there is CP violation ($\theta \neq 0$ and $\phi \neq 0$).

For $m_a = 10$~GeV, as a representative value for our study, we find that $\phi \lsim 10^{-7}$ if $C_{\tau e} = 1$, $\Lambda = 1$~TeV, and  $\theta$ is near $\pi/4$, in light of the experimental bound $|d_e| < 4.1 \times 10^{-30}\ e\ {\rm cm}$ \cite{Roussy:2022cmp}.  Hence, it is justified to assume $\phi=0$ for our analysis, given the very stringent limit from the electron EDM.  For a more detailed discussion, see Refs.~\cite{DiLuzio:2020oah,Bauer:2021mvw,DiLuzio:2023lmd}.

Notably, the sign of $\Delta a_e$ (Eq.~(\ref{eq:mdm})) is dependent on the value of the parity-violating angle $\theta$; the shift is negative for $\cos{2\theta} \gsim 0$ and positive for $\cos{2\theta} \lsim 0$ (up to factors of $m_e/m_\tau$). This is interesting in light of the most recent experimental determination of $a_e$, where the sign of the discrepancy with the SM depends on which experimental measurement of the fine-structure constant $\alpha$ is used. In particular, if one uses the measurement from rubidium $\alpha(\rm Rb)$ \cite{Morel:2020dww}, there is a $+ 2.2\sigma$ deviation
\begin{equation}
    \Delta a_e({\rm Rb}) = (34 \pm 16)\times 10^{-14},
\end{equation}
whereas if one uses the measurement from cesium $\alpha(\rm Cs)$ \cite{Parker:2018vye}, there is a $- 3.7\sigma$ deviation
\begin{equation}
\Delta a_e ({\rm Cs}) = (-101 \pm 27)\times 10^{-14}
\end{equation}
from the SM expectation.  A flavor-violating ALP interaction of the form considered here can explain an anomalous contribution to $\Delta a_e$ of either sign.

\begin{figure}[t!]
    \includegraphics[width = \linewidth]{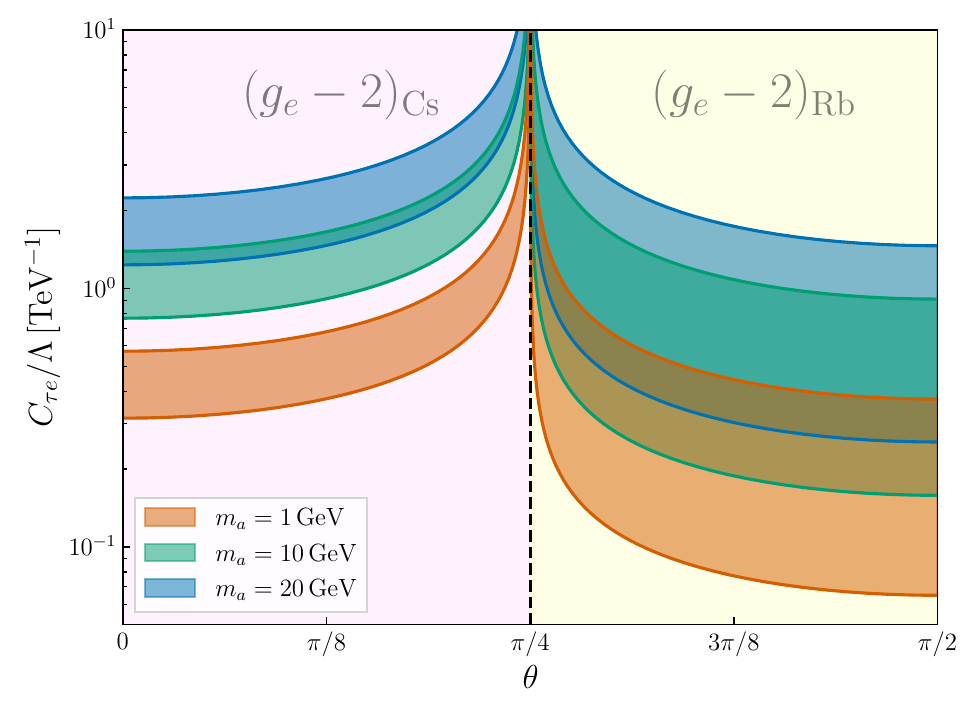}

    \caption{2$\sigma$ explanations for $\Delta a_e({\rm Cs)}$ (pink-shaded region, $\theta \lsim \pi/4$) and for $\Delta a_e({\rm Rb)}$ (yellow-shaded region, $\theta \gsim \pi/4$) at $m_a = 1$, $10$ and $20~{\rm GeV}$. Which anomaly is explained is dependent on the sign of $\Delta a_e$, which in turn is dependent on the parity-violating angle $\theta$. To good approximation, the sign flips at $\theta = \pi/4$, but there are $O(m_e/m_\tau)\sim 10^{-4}$ corrections. Digitized plot data for this and other figures are available as ancillary files on arXiv.org. \label{fig:g2e}}
\end{figure}

\begin{figure*}
    \centering
    \includegraphics[width = 0.9\linewidth]{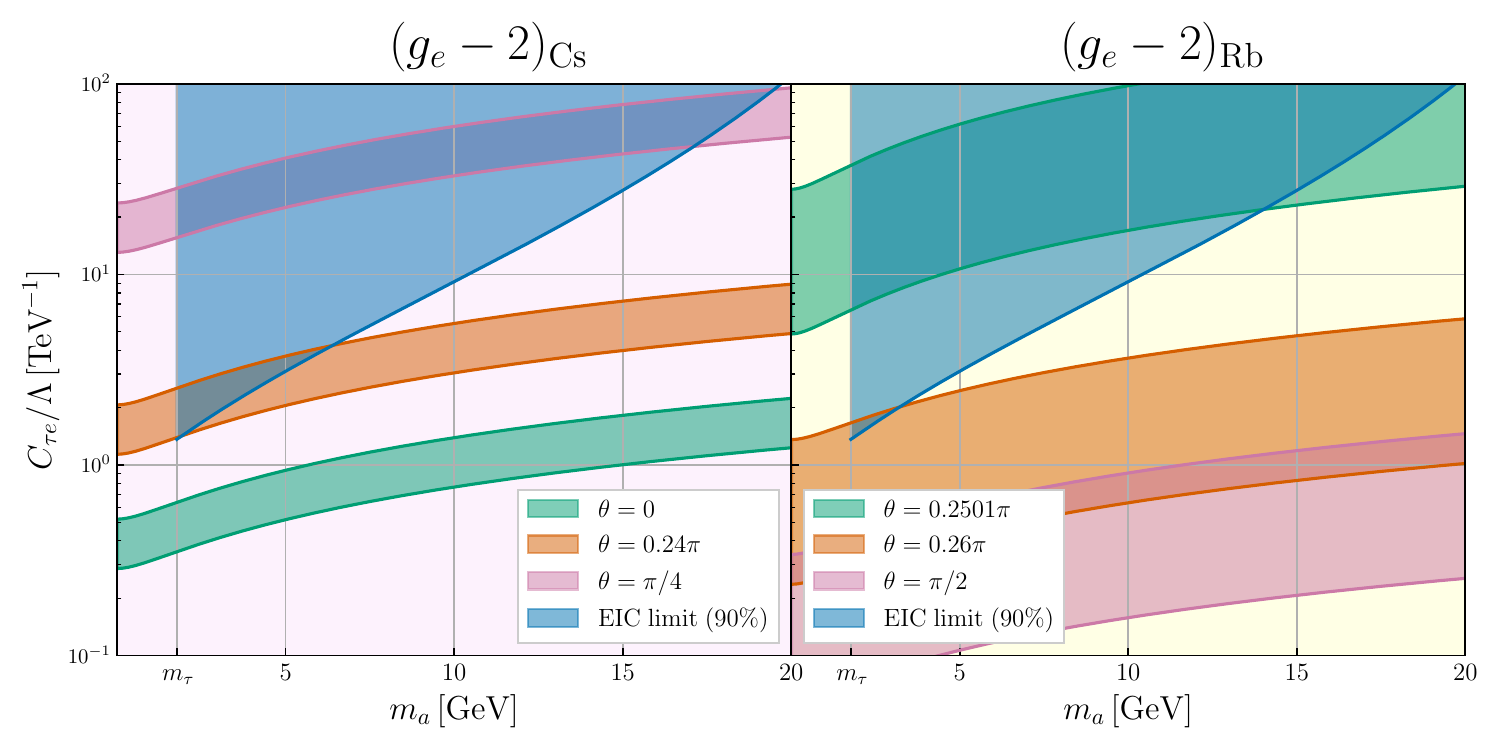}

    \caption{$2\sigma$ explanations for $\Delta a_e({\rm Cs})$ anomaly  (left, pink-shaded region) and $\Delta a_e({\rm Rb})$ anomaly  (right, yellow-shaded region).  Limits placed on $C_{e\tau}$ at the EIC at $90\%$ from Ref.~\cite{Davoudiasl:2021haa} are shown in blue. When the ALP $e$-$\tau$ coupling is chiral or near-chiral ($\theta \approx \pi/4$), the EIC can probe an explanation for either the cesium or rubidium anomaly.\label{fig:EIC_limits}}
    
\end{figure*}

\section{Results\label{sec:results}}

It is straightforward to compute the value of $C_{\tau e}$ that would explain one of these anomalies for a given angle $\theta$ and ALP mass $m_a$. The result is shown in Fig.~\ref{fig:g2e} for $m_a = 1$, $10$, and $20\,{\rm GeV}$ and $0 < \theta < \pi$. Explanations for $\Delta a_e ({\rm Rb})$ (positive) are in the yellow-shaded region whereas explanations for $\Delta a_e ({\rm Cs})$ (negative) are in the pink-shaded region.  As the figure shows, explanations for the $\Delta a_e$ anomaly can be found for $\mathcal{O}(1)$ or smaller $C_{\tau e}$ with $\Lambda \sim 1$ TeV over a wide range of ALP masses and PV angles. Notably, the magnitude of the coupling required to explain either anomaly grows for a near-chiral coupling ($|\theta| \rightarrow \pi/4$). 

\begin{figure}
    \centering
    \includegraphics[width = \linewidth]{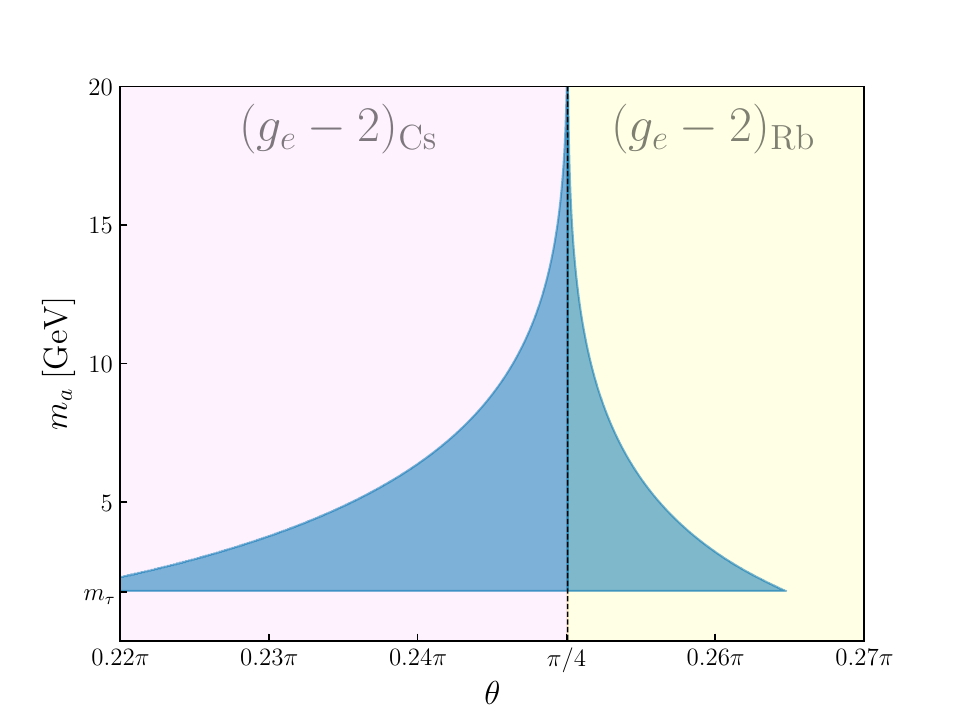}
    \caption{The masses $m_a$ and parity-violating angles $\theta$ for which the EIC can probe a $2\sigma$ explanation to the $g_e - 2$ anomaly using $\alpha({\rm Cs})$ (pink shaded region) and $\alpha({\rm Rb})$ (yellow shaded region). }
    \label{fig:EIC_explanation}
\end{figure}

As detailed in Ref.~\cite{Davoudiasl:2021mjy}, the EIC is in a unique position to probe the coupling $C_{\tau e}$, especially when the diagonal ALP couplings are relatively small. Chiral new physics is of particular interest at the EIC due to its polarization capabilities, and is well-motivated due to the asymmetry between left- and right-handed fields already present in the SM. For an example of a UV BSM setup which has ALPs with chiral LFV interactions, see Ref.~\cite{Davoudiasl:2017zws}. 

In Fig.~\ref{fig:EIC_limits}, we plot the $90\%$ confidence level projected exclusions on $C_{\tau e}$ from the EIC with ${\cal L} = (100/A)~{\rm fb}^{-1}$, discussed in   Ref.~\cite{Davoudiasl:2021haa}, and the corresponding explanations to the cesium anomaly (left, pink-shaded background) and the rubidium anomaly (right, yellow-shaded background). We see that the EIC is able to probe near-chiral ALP-lepton couplings which offer explanations to either $g_e-2$ anomaly. With a better $\tau$ efficiency and a dedicated background analysis to improve on the limits from Ref.~\cite{Davoudiasl:2021mjy}, the EIC may be able to probe a wider range of parity violating (PV) couplings.

If the PV angle is near-chiral, the EIC's polarization control of the electron beam may allow determination of the angle $\theta$. The EIC is expected to achieve over $70\%$ polarization of the electron beam \cite{AbdulKhalek:2021gbh}. Given production cross-sections dependent on the electron beam polarization $\sigma_L$ and $\sigma_R$, and a polarization fraction $p$, the left-right asymmetry observable at the EIC is
\begin{align}
    r_{LR} &= \frac{[p\sigma_L + (1-p)\sigma_R] - [(1-p)\sigma_L + p\sigma_R]}{\sigma_L + \sigma_R}\\
    &= (2p - 1)\frac{\sigma_L - \sigma_R}{\sigma_L + \sigma_R}.
\end{align}
For this ALP model with PV angle $\theta$, this corresponds to a left-right asymmetry
\begin{equation}
    r_{LR}(\theta) = (2p-1)\sin{2\theta}.
\end{equation}
This is extremal at $\theta = \pm\pi/4$.  If a flavor-violating ALP signal were to be discovered at the EIC, probing the left-right asymmetry of the signal to determine the PV angle $\theta$ would give an opportunity to directly verify the connection between such a signal and its contribution to $g_e - 2$.  The experimental reach of the EIC relevant for probing the $g_e - 2$ anomaly is shown as a function of the PV angle $\theta$ in Fig.~\ref{fig:EIC_explanation}.

\section{Concluding Remarks\label{sec:conclusions}}

Axion-like particles (ALPs) can arise in a variety of models and may, on general grounds, have flavor violating interactions.  In this work, we considered lepton flavor violation  and in particular $e-\tau$ transitions mediated by ALPs.  We also accounted for the possibility that ALPs can have parity violating interactions, akin to the electroweak sector of the SM.  With these assumptions, we found that ALPs can explain present hints for electron $g_e-2$ deviations whose sign depends on the choice of the precision value of the fine structure constant $\alpha$ that is used as input for the theory prediction.  

It is not yet clear which, if any, of the $g_e-2$ deviations would survive once the experimental extraction of $\alpha$ is disambiguated.  However, assuming that the size of the possible deviation remains at the current levels, we showed that the EIC can constrain the requisite $e-\tau$ coupling, for nearly chiral interactions and ALP masses above $m_\tau$, up to $\sim 20$~GeV.  Probing the smaller $e-\tau$ coupling which is needed to explain $g_e - 2$ deviations for PV angles $\theta$ away from $\pi/4$ will require improvements to the EIC search, or complementary searches in other future experiments for $e-\tau$ flavor-violating ALPs.

In addition, the $e$-beam polarization capabilities of the EIC can provide a unique probe of the parity violating angle that parameterizes departures from the purely chiral limit, over the above mass range.  We noted that ALP couplings can also include a CP violating phase, but the severe bounds on the electric dipole moment of the electron imply that such a phase must be extremely tiny, at the level of $\ord{10^{-7}}$ or less. 

Open questions, like the identity of dark matter, make a compelling case for new physics, yet allow a vast array of possibilities for where it may reside.  This motivates consideration of a broad set of ideas and the venues where they could be tested.  The capabilities of the EIC can lend themselves to searches for some of the signals that arise in well-motivated frameworks, like that of the ALP interactions  studied here; further work along this  direction hence seems warranted.

\section*{Acknowledgements}
We thank Marvin Schnubel and Sebastian Trojanowski for discussions.  
 This work is supported by the U.S. Department of Energy under Grant Contracts DE-SC0012704 (H.~D.) and DE-SC0010005 (E.~N. and R.~M.).  This material is based upon work supported by the U.S. Department of Energy, Office of Science, Office of Workforce Development for Teachers and Scientists, Office of Science Graduate Student Research (SCGSR) program. The SCGSR program is administered by the Oak Ridge Institute for Science and Education for the DOE under contract number DE‐SC0014664.

\begin{raggedright}
\bibliography{electron-gm2-eic}
\end{raggedright}

\end{document}